# Attention-Based Explainability for Structure–Property Relationships

Boris N. Slautin[1,*], Utkarsh Pratiush[2], Yongtao Liu[3], Hiroshi Funakubo[4], Vladimir V. Shvartsman[1], Doru C. Lupascu[1], Sergei V. Kalinin[2,5,*]

[1] *Institute for Materials Science and Center for Nanointegration Duisburg-Essen (CENIDE), University of Duisburg-Essen, Essen, 45141, Germany*

[2] *Department of Materials Science and Engineering, University of Tennessee, Knoxville, TN 37996, USA*

[3] *Center for Nanophase Materials Sciences, Oak Ridge National Laboratory, Oak Ridge, TN 37831, USA*

[4] *Department of Material Science and Engineering, Tokyo Institute of Technology, Yokohama 226-8502, Japan.*

[5] *Pacific Northwest National Laboratory, Richland, WA 99354, USA*

---

[*] Authors to whom correspondence should be addressed: boris.slautin@uni-due.de and sergei2@utk.edu

**Abstract**

Machine learning methods are emerging as a universal paradigm for constructing correlative structure-property relationships in materials science based on multimodal characterization. However, this necessitates development of methods for physical interpretability of the resulting correlative models. Here, we demonstrate the potential of attention-based neural networks for revealing structure–property relationships and the underlying physical mechanisms, using the ferroelectric properties of $PbTiO_3$ thin films as a case study. Through the analysis of attention scores, we disentangle the influence of distinct domain patterns on the polarization switching process. The attention-based Transformer model is explored both as a direct interpretability tool and as a surrogate for explaining representations learned via unsupervised machine learning, enabling the identification of physically grounded correlations. We compare attention-derived interpretability scores with classical SHapley Additive exPlanations (SHAP) analysis and show that, in contrast to applications in natural language processing, attention mechanisms in materials science exhibit high efficiency in highlighting meaningful structural features.

**I. Introduction**

Understanding structure-property relationships is a central challenge across materials science, chemistry, and molecular biology. In materials science, the arrangement of microstructural features can control mechanical strength, electrical conductivity, ferroelectric response, and electromechanical coupling.[1-6] In chemical systems, molecular architecture governs thermodynamic stability, reaction kinetics, and catalytic activity.[7-9] In biological contexts, biomolecular conformations dictate enzymatic function and binding specificity.[10-12] The identification of the structural elements responsible for promoting the specific functionalities of material behavior is pivotal in elucidating and understanding the underlying mechanisms that govern material properties.

An effective approach to probing structure–property relationships is the integration of imaging and spectroscopic techniques. Imaging methods, including optical microscopy, scanning probe microscopy (SPM), electron microscopy (EM) and others, enable rapid visualization of microstructural patterns.[13-18] Meanwhile, spectroscopic approaches, for instance, Raman spectroscopy,[19-21] electron energy loss spectroscopy (EELS),[22-27] or switching spectroscopy piezoresponse force microscopy (SS-PFM),[28-31] offer insights into local functional properties. The integration of spectroscopic and imaging techniques enables to capture both structural features and local functionalities, providing a foundation for uncovering structure-property relationships. However, leveraging such experimental datasets requires the construction of models that can relate structure to functionality. While physics-based models offer the highest generalizability and most interpretable insights, their development is often hindered by the heterogeneity of real materials and complexity of the image- and spectrum generation models, making them often infeasible to be constructed in practice. For example, over the past decades, significant progress has been made in modeling the local PFM response, including accounting for various parasitic contributions and the dynamics of polarization switching under the SPM probe.[31-34] However, in practice, the physical deconvolution of the measured spectra remains largely unresolved. In EELS spectroscopy, despite substantial advances in physics-based modeling, a fully predictive and quantitatively accurate description of experimental spectra remains a challenge.[35] Even for methods such as nanoindentation, full analysis of the measured elastic modulus and hardness in complex systems, such as multilayered thin films, requires accurate modeling of the phenomena including unloading-induced plasticity, heterogeneous stress distributions, layer undulation, and substrate effects, all of which hamper quantitative characterization.[36-39] Correspondingly, much attention has been attracted to exploring alternative approaches such as empirical correlation models.

During the last decades, advances in machine learning (ML) have been providing effective alternatives for handling complex, high-dimensional datasets.[40-44] The supervised regression models widely used in practice are suited for predicting scalar or low-dimensional vector properties from local structural descriptors.[45-49] In contrast, encoder–decoder architectures and variational autoencoders (VAE) excel supervised regression when both structure and functionality must be represented as complex, high-dimensional vectors.[50-58] These unsupervised deep learning frameworks learn latent representations that jointly capture structural motifs and their functional consequences, enabling the discovery of non-obvious patterns in large experimental or computational datasets. Deep Kernel Learning (DKL), which combines kernel-based Gaussian processes with deep neural networks, reduces the amount of data required for model training and facilitates a transition from purely statistical learning toward active learning strategies, which are widely employed in automated experimental workflows.[59, 60] In all cases, leveraging ML models to build correlative structure–property models raises the critical challenge of understanding the underlying physical mechanisms driving those correlations. Thus, we need reliable algorithms to enhance the explainability of the obtained relationships by ML models.

The need to improve the interpretability of ML models has given rise to eXplainable Artificial Intelligence (XAI).[61-63] XAI refers to a broad set of methods aimed at making complex ML models more understandable, turning their predictions into human-interpretable insights that can reveal the underlying physical mechanisms or governing principles.[64-66] These approaches can be applied both as a forensic tool for post-hoc (after training) and ante-hoc (intrinsic) analysis.[63, 66] Feature importance techniques, which estimate the contribution of each input descriptor to a prediction of a model, play a pivotal role in uncovering the physics behind nontrivial structure-property correlations in materials.[67-70] Over the past decade, multiple approaches of varying complexity have been proposed. Permutation importance estimates the importance of a feature by measuring the decrease in model performance when the values of that feature are randomly permuted, thereby breaking the relationship between the feature and the target.[71] Saliency map visualizations and similar techniques rely on computing the gradient of the output of the model with respect to its input features, defining which inputs have the strongest influence on the prediction.[72-74] Local Interpretable Model-Agnostic Explanations (LIME) focuses on interpreting individual prediction instances by approximating the complex model locally with a simpler, interpretable surrogate model.[75-77] One of the most widely adopted approaches for interpreting ML models in scientific research is SHAP (SHapley Additive exPlanations) method.[66, 78] SHAP is increasingly used in materials science to interpret

ML predictions across multiple scales spanning from identifying key molecular features in drug discovery and materials,[79-82] to optimizing structural descriptors in porous materials,[83-85] and guiding fabrication parameters in device-level applications like perovskite solar cells.[86, 87]

In addition to ad-hoc methods developed to enhance explainability, certain ML models inherently possess mechanisms for identifying the relevant correlations. The simplest example of such an approach is linear regression, where the coefficients directly indicate the importance. Following the path from the root to the leaves in a Decision Tree model reveals the decision-making logic and enables the assessment of feature importance. Among the deep learning models, Transformers possessing an intrinsic self-attention mechanism provide the opportunity for feature importance estimations.[88]

The attention mechanism assigns weights to the elements of the input descriptor, enabling the model to focus on the most important features rather than treating all input information equally.[88] This not only enhances the accuracy of model predictions, but also provides insights into the decision-making process. Importantly, several works have highlighted the limitations of using attention for model interpretability, as attention weights do not always align with feature importance or causal input influence.[89-91] However, the effectiveness of attention-based metrics in providing insights into feature importance and highlighting meaningful input structures has been demonstrated for various materials science problems. Louis *et al.* suggested a Global Attention Graph Neural Network (GAGNN) that enhances the graph neural network by the global attention layer.[92] This layer allows GAGNN to weight importance of different atomic interactions when predicting crystal properties. Wang *et al.* introduced a Transformer-like neural network (CrabNet) for predicting material properties directly from chemical composition. The self-attention mechanism model interactions between chemical elements enables interpretable insights into how each element contributes to the predicted material behavior.[93] Lu *et al.* introduced an attention-based convolutional neural network (Attention CNN) for predicting the properties of Heusler alloys.[94] They showed that the Attention CNN outperformed conventional CNN models and identified values for property prediction element-level information (e.g., electronegativity, atomic number). The efficacy of attention scores for estimating elemental contributions to the properties of inorganic materials was recently demonstrated in the hybrid CrysCo approach, which combines GNN with a Transformer-based attention mechanism.[95]

Here, we employ attention-based neural architectures and the SHAP method to identify the key structural motifs within input descriptors that govern local ferroelectric functionality. We benchmark the attention-based model against SHAP on a simple synthetic 1D dataset to assess its ability to highlight informative features. Then we apply both methods to experimental PFM data measured on a PbTiO₃ thin film, using the attention model both as a direct interpretability tool and as a surrogate to explain unsupervised ML representations, thereby revealing structure–property correlations and the underlying physical mechanisms. Overall, this work aims to assess the effectiveness of attention-based approaches in directly uncovering structure–property relationships from experimental data obtained via microscopic and spectroscopic techniques, and in improving the interpretability of unsupervised ML models applied in experimental materials science.

## II. Attention based explainability

### II.A. Background concept: attention in Transformer architecture.

The attention mechanism was originally introduced by Bahdanau *et al.*,[88] and subsequently popularized by Vaswani *et al.*,[96] to address key limitations of recurrent neural networks (RNNs) and CNNs in modeling tasks. Specifically, the attention algorithm was proposed to overcome the limited capacity of RNNs to capture long-range dependences and the low interpretability of CNNs due to their hierarchical feature extraction. Attention enabled the model to focus on the most relevant information in the input data while disregarding parts that have little influence on the target property. This not only significantly improves model performance but also facilitates interpretability.

The core idea of the attention mechanism is to highlight informative components by computing a weighted sum over a context sequence, defining a set from which relevant information is extracted, hence allowing models to directly focus on the relevant elements regardless of their position or distance. Importantly, in the general case, the context sequence is not necessarily the same as the input and can include information about the current state of the model, previous predictions, etc. Mathematically, in attention mechanisms, inputs are linearly projected into three distinct representations: queries ($Q$), keys ($K$), and values ($V$). The attention operation computes similarity scores between the queries and keys to weight the corresponding values by the dot-product defined as:

$$Attention(Q, K, V)_i = \sum_j softmax(\frac{Q_i K_j^T}{\sqrt{d_k}}) V_j \quad (1)$$

where $d_k$ is the dimension of the keys. The attention learning process involves optimizing the projection matrices that generate $Q$, $K$, and $V$ from the input data. These parameters are learned via backpropagation, enabling to assign higher weights to the most relevant elements and improve task-specific performance of the model.

The special case of the attention mechanism is **self-attention**. Here, $Q$, $K$, and $V$ are all derived from the same input sequence:

$$Q = XW_Q, K = XW_k, V = XW_V \qquad (2)$$

where $X$ is a set of input vectors; $W_Q$, $W_k$, and $W_V$ are the query, keys, and values matrices respectively. This allows each element in the sequence to relate to all others, enabling the model to capture contextual relationships and internal dependences across the entire input.

The self-attention mechanism is permutation-invariant by design, meaning it treats the input as an unordered set rather than a sequence, in contrast to RNNs and CNNs, which are inherently order-sensitive. To incorporate sequential information (e.g. spatial or temporal) **positional encoding** is used. The idea behind positional encoding is to add a position-dependent vector to each input embedding, enabling the model to distinguish the order of elements in the sequence. These encodings can be fixed (e.g. sinusoidal functions) or learned. In our implementation, we use a learnable positional embedding, which allows the model to discover task-specific spatial priors from the data. All model codes and raw results are available on GitHub (see Data Availability Statement).

**II.B. Implementation**

The Transformer models with the self-attention layer were employed for direct prediction of the target functionality by the input spectra (synthetic example) or small squared structural image patches representing local ferroelectric domain structures around measurement points. The regressor is designed to process 1D input vectors, such as synthetic spectra or flattened 2D structural patches. The model begins with a linear embedding layer that projects each scalar input to a 64-dimensional hidden space. A learnable positional encoding is added to preserve sequence order. The core of the model consists of a self-attention layer, followed by layer normalization and a feed-forward network with two linear layers (dimensions: 64–256–64) and Gaussian Error Linear Unit (GELU) activation. A second normalization layer follows the feed-forward block. After processing, the output is mean pooled along the sequence dimension and passes through a final linear layer to produce a scalar prediction. The Transformer models is trained using the Adam optimizer with a learning rate of 0.001, a batch size of 128, and for 200 epochs. The loss function used is mean squared error (MSE).

As a second approach, we applied Transformer as a surrogate model for the variational autoencoder. We employed a rotationally invariant rVAE as an unsupervised model to compress structural image patches into a low-dimensional latent space.[97, 98] The rVAE model were implemented by the PyroVED Python package.[99] The rVAE consists of an encoder and decoder, each composed of two fully connected layers with 128 neurons and ReLU activation. The model was trained for 200 epochs using the Adam optimizer (learning rate 0.001, batch size 128) with a latent space dimensionality of either 2 or 10. To interpret the VAE latent space, a Transformer regressor with the same architecture as described above was trained to predict target functionality directly from the VAE-generated latent vectors. The attention weights from this model were analyzed to attribute importance to individual latent variables and link them to underlying structural features in the material. In parallel, SHAP analysis was applied to provide an additional interpretability and benchmark attention approach.

To ensure the coverage of previous relevant literature, the standard literature search was complemented by the assistance from the large language model ChatGPT-4.0 (OpenAI) used in literature search for the introduction of this paper. The prompt and raw model outputs are provided in the supplementary materials.

**III. Direct explainability of the supervised models**

An attention-based Transformer model was used as a representative supervised model to evaluate the effectiveness of self-attention mechanisms and SHAP analysis in identifying the most relevant components within input descriptors. The analysis was first carried out on synthetic Gaussian spectra and then applied to real ferroelectric PFM data.

While self-attention weights define the pairwise relationships between all elements of the input descriptor, the attention matrix for an input of length $N$ has dimensions $N \times N$. To estimate the contribution of each element within the descriptor to prediction, we compute an *importance* vector, defined as the vector of average attention weights of each element relative to all others in the descriptor. Attention weights and therefore attention-based importance are always non-negative, reflecting the relative focus of the model on different elements within the input. In contrast, Shapley values, the core metric in SHAP analysis, can be either positive or negative, indicating whether a feature increases or decreases the prediction. Here, feature importance can be estimated by the absolute value of the Shapley values, representing how strongly each element contributes to the output, regardless of sign.

**III.A. Gaussian spectra analysis.**

As a simple synthetic example with known ground truth, we generated a dataset consisting of spectra composed of three Gaussian peaks with randomly sampled parameters.

To simulate experimental conditions, additive white noise was introduced to the generated spectra (Figure S1a). Let the synthetic spectrum $S(x)$ be defined as:

$$S(x) = \sum_{i=1}^{3} A_i \, exp\left(-\frac{x-\mu_i}{2\sigma^2}\right) + \varepsilon \qquad (4)$$

where $A_i \sim Uniform(0.5, 2.0)$, $\mu_i$ and $\sigma_i$ are sampled from uniform ranges for position and width, and $\varepsilon \sim Normal(0, 0.05)$ is Gaussian noise. The amplitude of the first peak was used as the target variable for the transformer regressor model.

The Transformer regressor was trained to predict the amplitude of the first of three peaks in the input spectra. Giving all the peaks are well-separated and do not overlap with each other, it is expected that only the region in the vicinity of the first peak position is valuable for the prediction. In this example the remnant part of the spectra plays the role of the excess information in the input data which is not needed for the target prediction.

The regressor trained on synthetic spectra achieved root mean squared error (RMSE) of approximately 0.05, corresponding to 4.24% of the average target amplitude on the 500 newly generated test spectra, which were used for the attention and SHAP analysis. Examples of absolute Shapley values and attention-based importance metrics for several individual spectra are shown in Figure S2. Both methods consistently highlight narrow regions around the location of the first spectral peak, demonstrating the ability of both approaches to identify informative features within high-dimensional input data (Figure 1a,b).

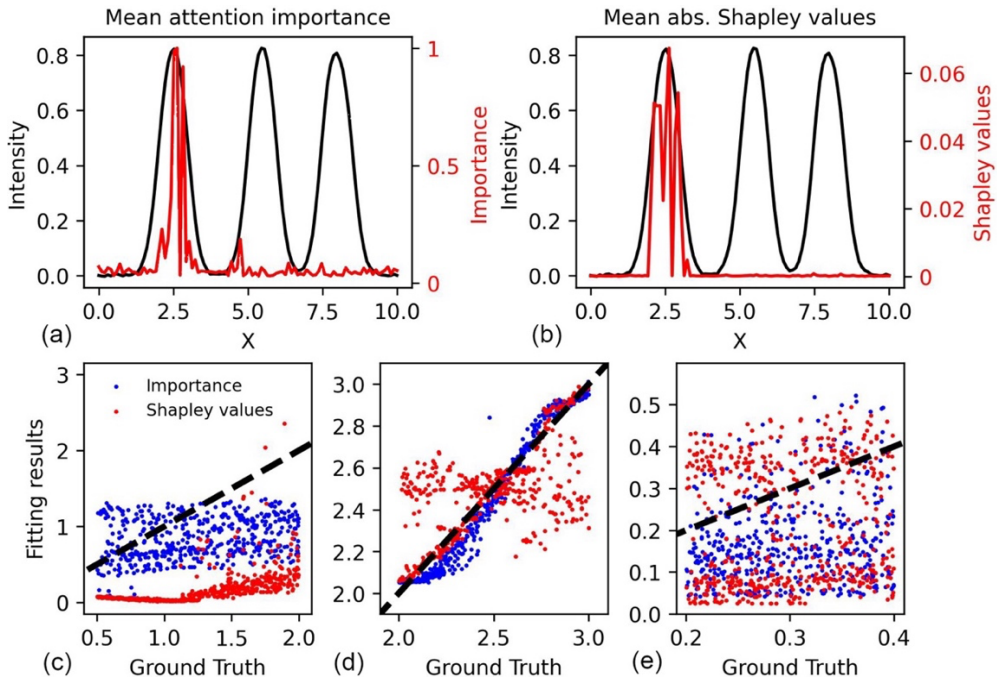

**Figure 1.** Dataset-averaged Gaussian spectra alongside corresponding (a) mean attention-based importance and (b) mean absolute Shapley values. Panels (c–e) show the dependence of fitted parameters, (c) amplitude, (d) peak location, and (e) width, on the ground truth values for both attention importance and Shapley values vectors.

To formalize the analysis described above, we fitted the absolute Shapley values and attention-based importance "spectra" using a single Gaussian peak. It is important to note that there is no physical or theoretical reason for these distributions to follow a Gaussian shape. However, this rough approximation allows for a simplified, quantitative characterization of their spread and localization to compare with the ground truth values.

The resulting fitting parameters including amplitude, peak location, and width were benchmarked against the ground truth values. A clear linear correlation was observed between the fitting parameters of the Shapley values and the ground truth peak amplitude, whereas no such correlation was found for the attention-based importance scores fitting. This discrepancy can be attributed to the underlying mechanisms of the two approaches. The amplitude of the Shapley values reflects the difference between the prediction and the average spectrum, making it sensitive to variations in peak amplitude. In contrast, attention weights primarily reflect positional relevance and are less influenced by absolute intensity, which may explain the absence of correlation with the peak amplitude.

We observed a clear correspondence between the fitted peak positions obtained from attention importance metrics and the ground truth peak locations. A similar trend was also found for the locations defined from Shapley values, although the linear dependence was less pronounced, reflecting their lower precision. The root mean squared error (RMSE) between the fitted and true peak positions was 0.08 (~3% of the average peak position value) for attention-based importance and 0.31 (~13%) for SHAP analysis. The corresponding Pearson correlation coefficients were 0.97 and 0.45, respectively. In contrast, no consistent relationship was found between the fitted and ground truth peak widths. We attribute this to the fact that the Gaussian width in the fitting procedure is primarily influenced by the overall quality of the fit, rather than by an interpretable feature of the input data. Overall, both algorithms have demonstrated their efficacy in the distinguishing principle for the prediction regions within the input descriptors.

**III.B Image-property regressor.**

Thick lead titanate ferroelectric films ($PbTiO_3$, PTO) grown by the H. Funakubo group were selected as a model ferroelectric system.[100] These films have been extensively characterized in prior studies, making them an ideal, well-understood benchmark for our analysis.[101-106] The local polarization maps were utilized for the attention-based framework benchmarking. These images were acquired by the Band-excitation piezoresponse force microscopy (BE PFM), yielding representative polarization maps (Figure S1b). The employed dataset consists of BE response effective "polarization" ($A\cos(\theta)$, $A$ – amplitude, $\theta$ – phase), cantilever resonant frequency ($f_{res}$), areas ($Area$), coercive voltages ($V_c$) and biases ($V_b$) of the local electromechanical loops measured at the same locations. It is important to note that, due to the specifics of the dataset collection, the values of $A$ and $\theta$ correspond to the average amplitude and phase derived from the measured loops. As a result, they do not fully represent the true PFM signal, but incorporate information related to the local domain structures suitable for the analysis.

The PFM $R\cos(\theta)$ map was employed as input (structural image) for the attention-regressor and cropped into the small rectangular patches each representing the local domain arrangement. The aim of the experiment is to forecast various ferroelectric-related functionalities based on the local domain arrangements captured within the image patches. The descriptor size, referring to the size of the image patches used as input, is a pivotal hyperparameter for the input data preprocessing. On one hand, increasing the patch size allows the model to incorporate more distant microstructural features that may influence the local response. However, excessively large patches introduce redundant information and noise, which can hinder precise prediction and reduce model performance. While the results presented below were primarily obtained using a descriptor size of 9×9 pixels (corresponding to ~100×100 nm²), results for other descriptor sizes can be found in the Supplementary Material.

We have trained an attention-based regressor to predict four parameters of the hysteresis loops by the patches: (1) the loop area ($Area$), representing the effective work of switching; (2) the coercive voltage ($V_c$), corresponding to the coercive field;[107] (3) the bias ($V_b$) which depends on configuration of local internal electric fields and (4) the contact resonance frequency, which reflects local variations in topography, electrostatic environment, and ferroelectric state as measured by the PFM signal. The SHAP analyses were also applied to the trained model. The aim of the experiment is to analyze variations and spatial redistributions of attention-based importance and Shapley values across different patches, to identify microstructural domain patterns most relevant for prediction and to align them with the corresponding target functionalities.

First, we analyzed the distributions of attention-based importance and Shapley values averaged across the image patch dataset (Figure 2). For the prediction of loop area and bias voltage, both attention and SHAP analyses revealed a clear prioritization of the central pixels within the descriptors (Figure 2j,l). A significantly weaker prioritization of the central region was observed for the coercive voltage (Figure 2k). For contact resonance frequency prediction, only the attention-based analysis shows a slightly higher average importance score in the central pixels, whereas the SHAP values exhibit an almost flat distribution (Figure 2i). This effect persists across different descriptor sizes, highlighting its physical origin rather than being an incidental artifact (Figure S3).

The observed patterns can be explained by the experimental geometry: the central positions within each image patch correspond to the probe location during scanning. Given the electric field distribution and the mechanical nature of the measured piezoresponse, prioritization of the central area naturally arises from the experimental setup. In contrast, elevated importance in non-central regions may reflect both the influence of neighboring structural elements (domain arrangement) and contributions from the non-local field. Such delocalization of the contact resonance frequency could arise from a combination of factors, including local topography, electrostatic interactions from the tip and cantilever, local elastic properties, and mechanical boundary conditions. In general, the similar trends observed in the average importance scores as a function of distance from the patch centers support the consistency between the two approaches.

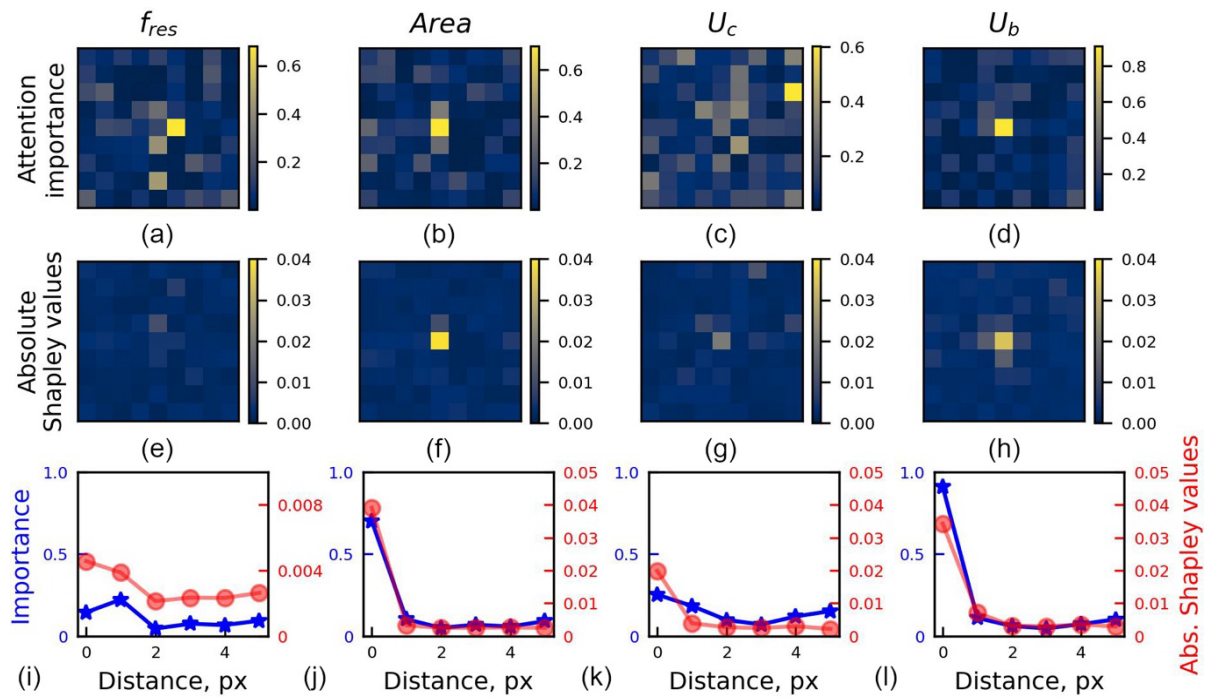

**Figure 2.** Average distributions of (a-d) attention-based importance and (e-h) absolute Shapley values for the prediction of contact resonance frequency, loop area, coercive voltage, and bias voltage, respectively, across structural image patches (patch size: 9×9 pixels). Panels (i-l) show the dependence of both importance metrics on the distance from the center of the patch.

The analysis of the averaged importance metrics distributions presented above enables insights into the characteristic length scales defining mechanisms underlying the formation of the target signals. However, the measured distributions represent a convolution of both microstructural effects and experimental geometry. Isolating the microstructural contribution from the geometric influences requires a study at the level of individual patches. We have analyzed the spatial distribution of the following two metrics defined for each image patch: 1) the number of "informative" pixels in each patch, where importance scores exceed 50% of the maximum value within that patch; and 2) the ratio of the importance at the central 3x3 pixels to the average importance across the remaining pixels in the patch (excluding the central ones). The former metric quantifies the degree of attention localization, while the ratio of the importance measures the extent to which the model prioritizes the central region within each patch.

For the resonance frequency predictor, we observed an elevated number of important pixels along the a–c domain walls (Figure 3a). A similar enhancement was also noticed in the importance ratio metrics (Figure 3e). In contrast, a distinct decrease in the importance ratio was found at the c–c domain walls. This observation is consistent with the underlying physics: at a–c domain walls, the elastic properties change markedly due to the rotation of the elastic tensor between domains, which strongly influences the cantilever resonance and related signals. At c–c domain walls, however, the elastic contrast is minimal, leading to negligible effects.

The importance ratio metrics for both the loop area and coercive voltage predictors show similar behavior, with values generally decreasing when the probe is positioned at domain walls (Figure 3f,g). This suggests that, at domain walls, the algorithm relies primarily on the surrounding domain configuration, reducing the relative importance of the response from the walls themselves. For both predictors, the distribution predominantly follows the a-domain pattern. However, slight differences in the metric values between the between regions of c-domains with opposite polarization directions can also be observed. The analysis of informative pixel distributions in structural images for these targets is limited by high noise levels, particularly in the case of the coercive voltage (Figure 3b,c).

The c-domain configuration plays a significant role in predicting the loop bias, as variations in both metrics clearly trace the shapes of regions with opposite c-domain polarization directions. (Figure 3d,h). This correlation can be explained by the imprint effect: built-in electric fields or asymmetric charge and defect distributions differ between c-domains with opposite polarization, leading to a domain-dependent shift of the hysteresis loop bias. Additionally, a pronounced increase in the importance ratio is observed at domain walls.

Similar spatial distributions of both metrics for these targets were also observed across other descriptor sizes (Figures S3, S4).

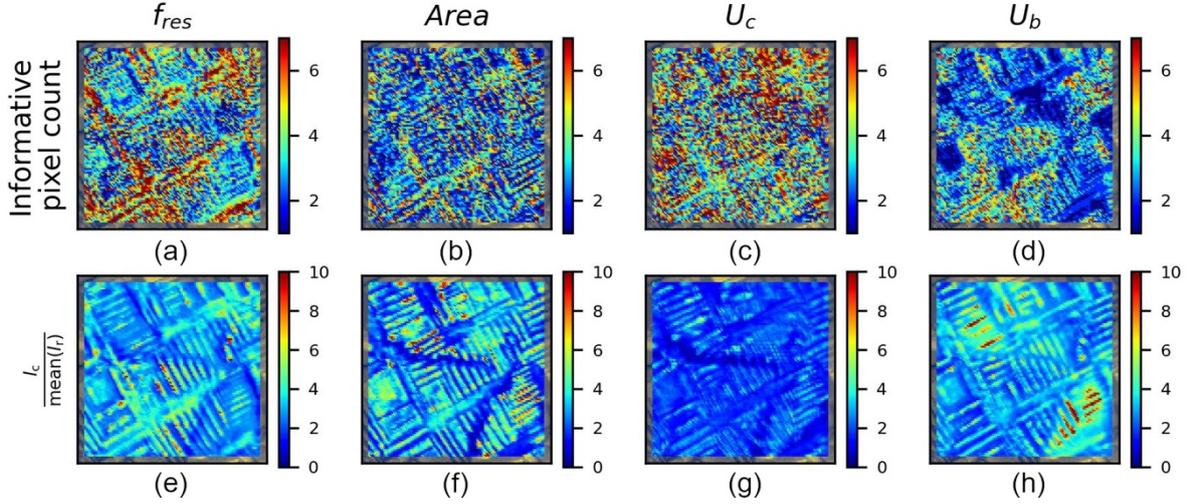

**Figure 3.** (a–d) Spatial maps showing the number of informative pixels within each image patch, defined as pixels with attention importance values greater than 0.5, for frequency, loop area, coercive voltage and loop bias prediction, respectively. (f–h) Corresponding maps of the ratio between the average attention importance of the central 3x3 pixels and the importance of the remaining pixels in the patch.

**IV. Interpretability in the latent space.**

Unsupervised VAE models can uncover hidden interdependences within high-dimensional datasets and mapping them into a lower-dimensional latent space in the form of latent variables. Additionally, the generative nature of VAEs makes them indispensable not only for studying structure–property relationships but also for guiding materials design.[51, 108-110] However, due to their inherently unsupervised nature, directly interpreting the properties encoded in the latent variables remains a significant challenge.

Patches of the structural image, representing local domain arrangements, were encoded into a 2D latent space using a rVAE. The isolation of the rotation angle into a single latent variable allows the structural influence to be deconvoluted from those arising due to the experimental geometry. The non-random distribution of target functionality values within this latent space highlights their dependence on the local domain structure (Figure 4a-d). However, these distributions are complex and nontrivial. An exception to this is the distribution of bias voltage, which shows a clear correlation with $z_2$ latent (Figure 4d). An attention-based regressor was trained to predict target functionalities from the corresponding latent vectors, and the distribution of attention-based importance metrics was analyzed within the latent space.

Additionally, SHAP analysis was applied to further interpret the contribution of individual latent variables.

Both the attention and SHAP algorithms prioritize different non-random regions within the latent space (Figure 4f,h,j,l), corresponding to distinct microstructural features (Figure 4e,g,i,k). The attention-based importance displays a more stepwise distribution, whereas Shapley values vary more gradually. We attribute this behavior to the low dimensionality of the input descriptor, which consists of just three elements: the angle and two latent variables. This limited input size makes the attention mechanism, originally designed for longer sequences, less suitable in this context. Alternatively, relatively primitive or coarse patterns of attention importance across the VAE latent space, may reflects the simplicity of the underlying feature representation.

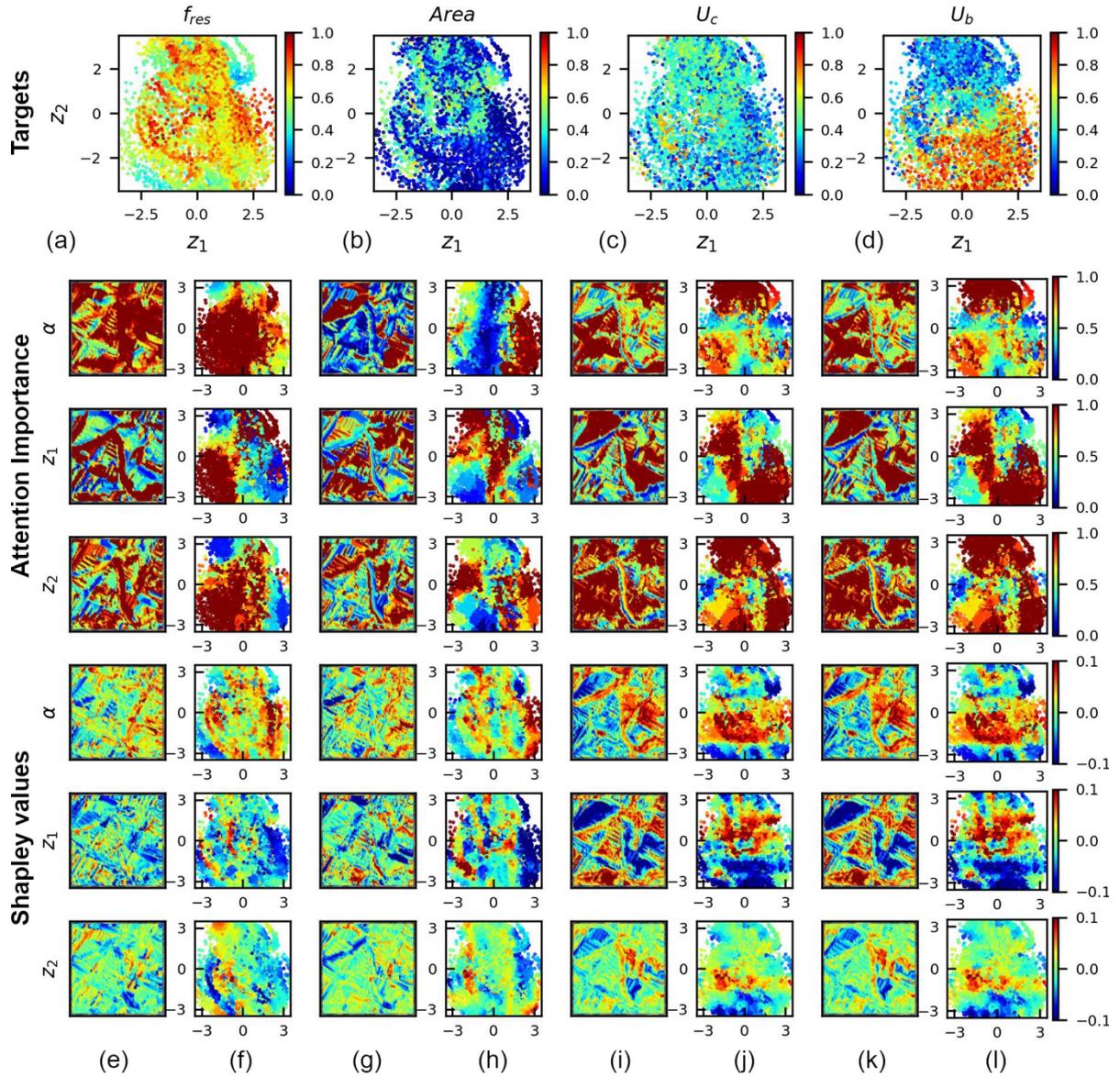

**Figure 4.** Distributions of the target properties within the 2D rVAE latent space: (a) contact resonance frequency, (b) loop area, (c) coercive voltage, and (d) bias voltage. Attention-based importance and Shapley values projected into both real space (e,g,i,l) and rVAE latent space (f,h,j,l) for the corresponding targets.

For further analysis, considering the diverse and complex domain structure of the PTO thin film, the structural image patches were encoded into a 10-dimensional rVAE latent space. If the angular parameter describing domain orientation is included, the system becomes effectively 11-dimensional, with the rVAE isolating this factor into a distinct latent variable. While direct visualization of such a high-dimensional latent space is not feasible, projecting these latent variables into real space reveals that the algorithms prioritize distinct structural features for different variables (Figure S6). The spatial distributions of importance scores correlate primarily with the *a*-domain structure, revealing its dominant influence on ferroelectric switching behavior and the contact resonance frequency.

Certain latent variables exhibit long-range spatial correlation, while others are characterized by much shorter correlation lengths (Figure 5a). We attribute variables with short correlation lengths to local features of densely packed *a-domains*, which tend to vary rapidly within the structural image. Oppositely, latent variables with long correlation lengths reflect presumable polarization direction change in *c-domains*. To estimate the contribution of each latent variable to the prediction of target functionalities, we computed the average attention importance and absolute Shapley values across the dataset (Figure 5b–d). The observed correlation between attention-based importance and absolutely Shapley value scores for multiple target properties and descriptor sizes, supporting the consistency of both interpretability approaches (Figure S7).

For bias voltage prediction, the most influential latent variables, which are identified consistently by both attention and SHAP analyses, were those with the longest spatial correlation lengths, primarily associated with the polarization direction distribution in c-domain pattern (Figure 5e). This indicates that bias voltage is largely governed by the c-domain arrangement, consistent with the results obtained from the direct image-to-property predictor. A similar, but less pronounced, relationship was observed in the absolute Shapley values for loop area prediction, suggesting that c-domain structures also contribute to the switching behavior (Figure 5c). In contrast, no clear correlation was found between the spatial correlation lengths of latent variables and their importance scores for the contact resonance frequency or coercive voltage (Figure 5b,e). The nearly uniform distribution of importance metrics across the latent variables may indicate either a comparable significance of the c- and a-domain arrangements, or a dominant influence of the a-domains. Unlike the c-domain features, which can be clearly associated with specific latent variables, the characteristics of the densely packed a-domain structure are dispersed across multiple latent variables. This dispersion causes the attention metrics to appear "blurred," making detailed analysis more challenging. Nevertheless, the reduced importance of the polarization direction in c-domains in comparison with loop bias for these signals is evident.

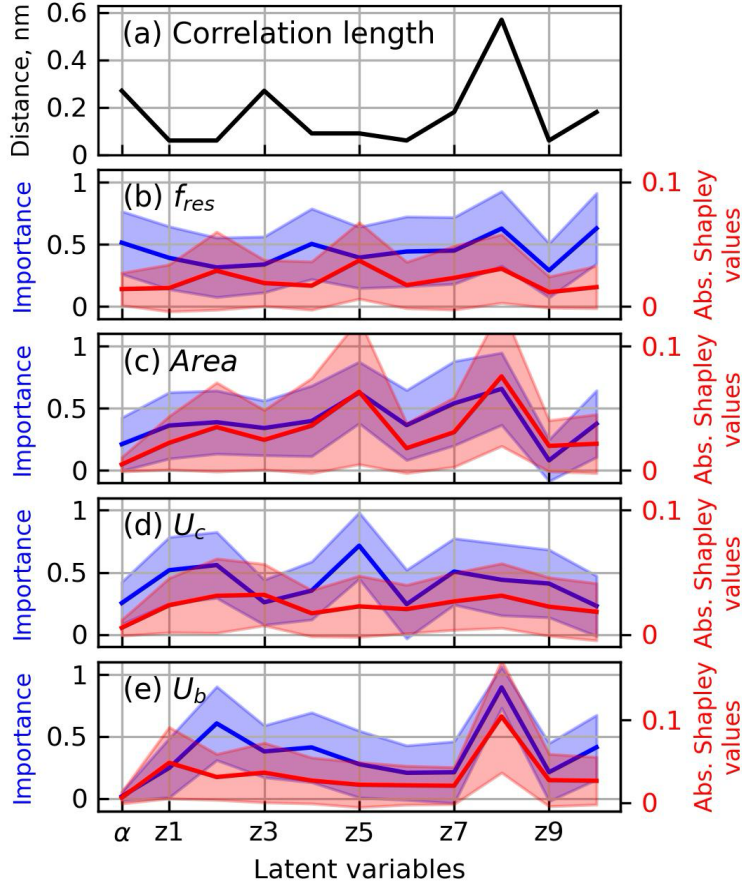

**Figure 5.** (a) Correlation length of the rVAE latent variables and (b-e) their corresponding average attention-based importance and absolute Shapley values for predicting contact resonance frequency, loop area, coercive voltage, and loop bias.

Summarizing the SHAP and attention-based analyses of both the rVAE latent space and the image-to-property predictor for the PTO dataset, we found that the hysteresis loop area and the coercive voltage are primarily determined by the configuration of *a-domains* in the region surrounding the tip. The decisive role of *a-domains*, and more specific a-c domain walls, is also evident in contact resonance frequency prediction, where the influence of *c-c domain walls* appears negligible. In contrast, for loop bias, which is strongly affected by built-in electric fields, the initial c-domain arrangement plays the dominant role.

## V. Summary

In this work, we demonstrated the potential of attention-based neural networks and SHAP analysis as complementary interpretability tools for understanding structure–property relationships in ferroelectric materials. Using both synthetic spectral data and real PFM measurements, we revised the ability of the attention approach to identify the most informative structural patterns within high-dimensional structural descriptors.

Although the specific target properties examined in this study primarily serve to demonstrate the methodology, the results indicate that the attention-based framework is broadly applicable to diverse material systems and functional responses. In contrast to applications in natural language processing,[89-91] we observe a notably high efficiency of attention-based scores for interpreting ML models in materials science, although the fundamental limitations highlighted in those works remain applicable. This can be attributed to the simpler architecture of the employed models and the more direct, physics-based relationship between structural features and material properties. The presented PFM examples exhibit increased complexity due to the intertwined effects of experimental geometry and underlying physical mechanisms. Interpreting the ML predictions therefore requires careful analysis, relying more on statistically robust trends than on individual predictions. A more direct and physically transparent interpretation is anticipated when spectroscopic input data are utilized.

We believe the attention-based models offer a promising pathway toward addressing one of the central challenges in ML-powered materials science: the interpretability of models. By bridging black-box predictions with physically meaningful features, these approaches enhance transparency and foster deriving scientific insight from ML-based predictions.

**Acknowledgments**

This work is supported (development and implementation, BNS and SVK) by NSF DMR proposal Award Number:2523284. The work was partially supported (UP) AI Tennessee Initiative at University of Tennessee Knoxville (UTK). This work (H.F) was also supported by Japan Science and Technology Agency (JST) as part of Adopting Sustainable Partnerships for Innovative Research Ecosystem (ASPIRE)(JPMJAP2312), MEXT Initiative to Establish Next-generation Novel Integrated Circuits Centers (X-NICS) (JPJ011438) and by MEXT Program: Data Creation and Utilization Type, Material Research and Development Project ( JPMXP1122683430). This work (PFM measurements) was supported by the Center for Nanophase Materials Sciences (CNMS), which is a US Department of Energy, Office of Science User Facility at Oak Ridge National Laboratory.

**Author Contributions**

**BNS**: Conceptualization; Investigation; Software; Writing – original draft. **UP**: Conceptualization; Software; Writing – original draft. **YL**: Data curation; Writing – review & editing. **HF**: Resources. **VVS:** Writing – review & editing. **DCL**: Writing – review & editing. **SVK**: Conceptualization; Supervision; Writing – review & editing.

**Data Availability Statement**

The analysis codes that support the findings of this study are available at https://github.com/Slautin/2025_Attention

**References**

(1) Clemens, H.; Mayer, S.; Scheu, C. Microstructure and Properties of Engineering Materials. *Neutrons and Synchrotron Radiation in Engineering Materials Science*; Staron P., Schreyer A., Clemens H., Mayer S., Ed.; Wiley, 2017; pp 1-20. DOI: 10.1002/9783527684489.ch1

(2) Barrett, C. R.; Nix, W. D.; Tetelman, A. S. *The principles of engineering materials*; Prentice hall, 1973.

(3) Arlt, G. The influence of microstructure on the properties of ferroelectric ceramics. *Ferroelectrics* 1990, *104* (1), 217-227. DOI: 10.1080/00150199008223825.

(4) Damjanovic, D. Ferroelectric, dielectric and piezoelectric properties of ferroelectric thin films and ceramics. *Reports on progress in physics* 1998, *61* (9), 1267.

(5) Li, F.; Zhang, S.; Damjanovic, D.; Chen, L. Q.; Shrout, T. R. Local structural heterogeneity and electromechanical responses of ferroelectrics: learning from relaxor ferroelectrics. *Advanced Functional Materials* 2018, *28* (37), 1801504. DOI: 10.1002/adfm.201801504.

(6) Gleiter, H. Nanostructured materials: basic concepts and microstructure. *Acta Materialia* 2000, *48* (1), 1-29. DOI: 10.1016/S1359-6454(99)00285-2.

(7) Avinash, M. B.; Govindaraju, T. Architectonics: Design of Molecular Architecture for Functional Applications. *Accounts of Chemical Research* 2018, *51* (2), 414-426. DOI: 10.1021/acs.accounts.7b00434.

(8) Wegener, S. L.; Marks, T. J.; Stair, P. C. Design Strategies for the Molecular Level Synthesis of Supported Catalysts. *Accounts of Chemical Research* 2012, *45* (2), 206-214. DOI: 10.1021/ar2001342.

(9) Wan, C.; Duan, X.; Huang, Y. Molecular Design of Single-Atom Catalysts for Oxygen Reduction Reaction. *Advanced Energy Materials* 2020, *10* (14), 1903815. DOI: 10.1002/aenm.201903815.

(10) Boehr, D. D.; Nussinov, R.; Wright, P. E. The role of dynamic conformational ensembles in biomolecular recognition. *Nature chemical biology* 2009, *5* (11), 789-796. DOI: 10.1038/nchembio.232.

(11) Huggins, D. J.; Biggin, P. C.; Dämgen, M. A.; Essex, J. W.; Harris, S. A.; Henchman, R. H.; Khalid, S.; Kuzmanic, A.; Laughton, C. A.; Michel, J. Biomolecular simulations: From dynamics and mechanisms to computational assays of biological activity. *Wiley Interdisciplinary Reviews: Computational Molecular Science* 2019, *9* (3), e1393. DOI: 10.1002/wcms.1393.

(12) Maria-Solano, M. A.; Serrano-Hervás, E.; Romero-Rivera, A.; Iglesias-Fernández, J.; Osuna, S. Role of conformational dynamics in the evolution of novel enzyme function. *Chemical Communications* 2018, *54* (50), 6622-6634. DOI: 10.1039/C8CC02426J.

(13) Reitzig, S.; Rüsing, M.; Zhao, J.; Kirbus, B.; Mookherjea, S.; Eng, L. M. "Seeing Is Believing"—In-Depth Analysis by Co-Imaging of Periodically-Poled X-Cut Lithium Niobate Thin Films. *Crystals* 2021, *11* (3), 288. DOI: 10.3390/cryst11030288

(14) Shur, V. Ya.; Shikhova, V. A.; Ievlev, A. V.; Zelenovskiy, P. S.; Neradovskiy, M. M.; Pelegov, D. V.; Ivleva, L. I. Nanodomain structures formation during polarization reversal in uniform electric field in strontium barium niobate single crystals. *Journal of Applied Physics* 2012, *112* (6). DOI: 10.1063/1.4754511

(15) Sheng, Y.; Best, A.; Butt, H.-J.; Krolikowski, W.; Arie, A.; Koynov, K. Three-dimensional ferroelectric domain visualization by Čerenkov-type second harmonic generation. *Opt. Express* 2010, *18* (16), 16539-16545. DOI: 10.1364/OE.18.016539.

(16) Nelson, C. T.; Winchester, B.; Zhang, Y.; Kim, S.-J.; Melville, A.; Adamo, C.; Folkman, C. M.; Baek, S.-H.; Eom, C.-B.; Schlom, D. G.; et al. Spontaneous Vortex Nanodomain Arrays at Ferroelectric Heterointerfaces. *Nano Letters* 2011, *11* (2), 828-834. DOI: 10.1021/nl1041808.

(17) Ushakov, A. D.; Hu, Q.; Liu, X.; Xu, Z.; Wei, X.; Shur, V. Ya. Domain structure evolution during alternating current poling and its influence on the piezoelectric properties in [001]-cut rhombohedral PIN-PMN-PT single crystals. *Applied Physics Letters* 2021, *118*, 232901. DOI: 10.1063/5.0055127.

(18) Shur, V. Ya.; Zelenovskiy, P. Micro-and nanodomain imaging in uniaxial ferroelectrics: Joint application of optical, confocal Raman, and piezoelectric force microscopy. *Journal of Applied Physics* 2014, *116* (6), 066802. DOI: 10.1063/1.4891397.

(19) Baddour-Hadjean, R.; Pereira-Ramos, J.-P. Raman Microspectrometry Applied to the Study of Electrode Materials for Lithium Batteries. *Chemical Reviews* 2010, *110* (3), 1278-1319. DOI: 10.1021/cr800344k.

(20) Krylov, A.; Krylova, S.; Kopyl, S.; Krylov, A.; Salehli, F.; Zelenovskiy, P.; Vtyurin, A.; Kholkin, A. Raman Spectra of Diphenylalanine Microtubes: Polarisation and Temperature Effects. *Crystals* 2020, *10* (3), 224. DOI: 10.3390/cryst10030224.

(21) Pistor, P.; Ruiz, A.; Cabot, A.; Izquierdo-Roca, V. Advanced Raman Spectroscopy of Methylammonium Lead Iodide: Development of a Non-destructive Characterisation Methodology. *Scientific Reports* 2016, *6* (1), 35973. DOI: 10.1038/srep35973.

(22) García de Abajo, F. J. Optical excitations in electron microscopy. *Reviews of Modern Physics* 2010, *82* (1), 209-275. DOI: 10.1103/RevModPhys.82.209.

(23) Koh, A. L.; Bao, K.; Khan, I.; Smith, W. E.; Kothleitner, G.; Nordlander, P.; Maier, S. A.; McComb, D. W. Electron Energy-Loss Spectroscopy (EELS) of Surface Plasmons in Single Silver Nanoparticles and Dimers: Influence of Beam Damage and Mapping of Dark Modes. *ACS Nano* 2009, *3* (10), 3015-3022. DOI: 10.1021/nn900922z.

(24) Kim, Y.-J.; Palmer, L. D.; Lee, W.; Heller, N. J.; Cushing, S. K. Using electron energy-loss spectroscopy to measure nanoscale electronic and vibrational dynamics in a TEM. *The Journal of Chemical Physics* 2023, *159*, 050901. DOI: 10.1063/5.0147356.

(25) Zhou, W.; Lee, J.; Nanda, J.; Pantelides, S. T.; Pennycook, S. J.; Idrobo, J.-C. Atomically localized plasmon enhancement in monolayer graphene. *Nature Nanotechnology* 2012, *7* (3), 161-165. DOI: 10.1038/nnano.2011.252.

(26) Yu, L.; Li, M.; Wen, J.; Amine, K.; Lu, J. (S)TEM-EELS as an advanced characterization technique for lithium-ion batteries. *Materials Chemistry Frontiers* 2021, *5*, 5186-5193. DOI: 10.1039/D1QM00275A.

(27) Qu, J.; Sui, M.; Li, R. Recent advances in in-situ transmission electron microscopy techniques for heterogeneous catalysis. *iScience* 2023, *26* (7), 107072. DOI: 10.1016/j.isci.2023.107072.

(28) Alikin, D.; Abramov, A.; Turygin, A.; Ievlev, A.; Pryakhina, V.; Karpinsky, D.; Hu, Q.; Jin, L.; Shur, V.; Tselev, A.; et al. Exploring Charged Defects in Ferroelectrics by the Switching Spectroscopy Piezoresponse Force Microscopy. *Small Methods* 2022, *6* (2), 2101289. DOI: 10.1002/smtd.202101289.

(29) Hong, S.; Woo, J.; Shin, H.; Jeon, J. U.; Pak, Y. E.; Colla, E. L.; Setter, N.; Kim, E.; No, K. Principle of ferroelectric domain imaging using atomic force microscope. *Journal of Applied Physics* 2001, *89*, 1377-1386. DOI: 10.1063/1.1331654.

(30) Jesse, S.; Baddorf, A. P.; Kalinin, S. V. Switching spectroscopy piezoresponse force microscopy of ferroelectric materials. *Applied Physics Letters* 2006, *88*, 062908. DOI: 10.1063/1.2172216.

(31) Kalinin, S. V.; Morozovska, A. N.; Chen, L. Q.; Rodriguez, B. J. Local polarization dynamics in ferroelectric materials. *Reports on Progress in Physics* 2010, *73* (5), 056502. DOI: 10.1088/0034-4885/73/5/056502.

(32) Yudin, P. V.; Hrebtov, M. Y.; Dejneka, A.; McGilly, L. J. Modeling the Motion of Ferroelectric Domain Walls with the Classical Stefan Problem. *Physical Review Applied* 2020, *13* (1), 014006. DOI: 10.1103/PhysRevApplied.13.014006.

(33) Brugère, A.; Gidon, S.; Gautier, B. Finite element method simulation of the domain growth kinetics in single-crystal $LiTaO_3$: Role of surface conductivity. *Journal of Applied Physics* 2011, *110*, 052016. DOI: 10.1063/1.3623762.

(34) Chen, J.; Gruverman, A.; Morozovska, A. N.; Valanoor, N. Sub-critical field domain reversal in epitaxial ferroelectric films. *Journal of Applied Physics* 2014, *116* (12), 124109. DOI: 10.1063/1.4896730.

(35) Colliex, C. From early to present and future achievements of EELS in the TEM. *Eur. Phys. J. Appl. Phys.* 2022, *97*, 38. DOI: 10.1051/epjap/2022220012.

(36) Tang, G.; Shen, Y. L.; Singh, D. R. P.; Chawla, N. Indentation behavior of metal–ceramic multilayers at the nanoscale: Numerical analysis and experimental verification. *Acta Materialia* 2010, *58* (6), 2033-2044. DOI: 10.1016/j.actamat.2009.11.046.

(37) Jamison, R. D.; Shen, Y. L. Indentation behavior of multilayered thin films: Effects of layer undulation. *Thin Solid Films* 2014, *570*, 235-242. DOI: 10.1016/j.tsf.2014.04.023.

(38) Clifford, C. A.; Seah, M. P. Modelling of nanomechanical nanoindentation measurements using an AFM or nanoindenter for compliant layers on stiffer substrates. *Nanotechnology* 2006, *17* (21), 5283. DOI: 10.1088/0957-4484/17/21/001.

(39) Phani, P. S.; Oliver, W. C.; Pharr, G. M. Understanding and modeling plasticity error during nanoindentation with continuous stiffness measurement. *Materials & Design* 2020, *194*, 108923. DOI10.1016/j.matdes.2020.108923.

(40) Schmidt, J.; Marques, M. R. G.; Botti, S.; Marques, M. A. L. Recent advances and applications of machine learning in solid-state materials science. *Npj Comput Mater* 2019, *5*, 83. DOI: 10.1038/s41524-019-0221-0.

(41) Malica, C.; Novoselov, K. S.; Barnard, A. S.; Kalinin, S. V.; Spurgeon, S. R.; Reuter, K.; Alducin, M.; Deringer, V. L.; Csányi, G.; Marzari, N.; et al. Artificial intelligence for advanced functional materials: exploring current and future directions. *Journal of Physics: Materials* 2025, *8* (2), 021001. DOI: 10.1088/2515-7639/adc29d.

(42) Butler, K. T.; Davies, D. W.; Cartwright, H.; Isayev, O.; Walsh, A. Machine learning for molecular and materials science. *Nature* 2018, *559* (7715), 547-555. DOI: 10.1038/s41586-018-0337-2.

(43) George, J.; Hautier, G. Chemist versus Machine: Traditional Knowledge versus Machine Learning Techniques. *Trends in Chemistry* 2021, *3* (2), 86-95. DOI: 10.1016/j.trechm.2020.10.007.

(44) Himanen, L.; Geurts, A.; Foster, A. S.; Rinke, P. Data-Driven Materials Science: Status, Challenges, and Perspectives. *Advanced Science* 2019, *6* (21), 1900808. DOI: 10.1002/advs.201900808.

(45) Agrawal, A.; Choudhary, A. Deep materials informatics: Applications of deep learning in materials science. *MRS Communications* 2019, *9* (3), 779-792. DOI: 10.1557/mrc.2019.73.

(46) Jha, D.; Ward, L.; Paul, A.; Liao, W.-k.; Choudhary, A.; Wolverton, C.; Agrawal, A. ElemNet: Deep Learning the Chemistry of Materials From Only Elemental Composition. *Scientific Reports* 2018, *8* (1), 17593. DOI: 10.1038/s41598-018-35934-y.

(47) Xie, T.; Grossman, J. C. Crystal Graph Convolutional Neural Networks for an Accurate and Interpretable Prediction of Material Properties. *Physical Review Letters* 2018, *120* (14), 145301. DOI: 10.1103/PhysRevLett.120.145301.

(48) Sandfort, F.; Strieth-Kalthoff, F.; Kühnemund, M.; Beecks, C.; Glorius, F. A Structure-Based Platform for Predicting Chemical Reactivity. *Chem* 2020, *6* (6), 1379-1390. DOI: 10.1016/j.chempr.2020.02.017.

(49) Roccapriore, K. M.; Kalinin, S. V.; Ziatdinov, M. Physics Discovery in Nanoplasmonic Systems via Autonomous Experiments in Scanning Transmission Electron Microscopy. *Advanced Science* 2022, *9* (36), 2203422. DOI: 10.1002/advs.202203422.

(50) Wei, R. Q.; Mahmood, A. Recent Advances in Variational Autoencoders With Representation Learning for Biomedical Informatics: A Survey. *Ieee Access* 2021, *9*, 4939-4956. DOI: 10.1109/Access.2020.3048309.

(51) Gómez-Bombarelli, R.; Wei, J. N.; Duvenaud, D.; Hernández-Lobato, J. M.; Sánchez-Lengeling, B.; Sheberla, D.; Aguilera-Iparraguirre, J.; Hirzel, T. D.; Adams, R. P.; Aspuru-Guzik, A. Automatic Chemical Design Using a Data-Driven Continuous Representation of Molecules. *ACS Central Science* 2018, *4* (2), 268-276. DOI: 10.1021/acscentsci.7b00572.

(52) Kalinin, S. V.; Dyck, O.; Jesse, S.; Ziatdinov, M. Exploring order parameters and dynamic processes in disordered systems via variational autoencoders. *Science Advances* 2021, *7* (17), eabd5084. DOI: doi:10.1126/sciadv.abd5084.

(53) Ziatdinov, M. A.; Yaman, M. Y.; Liu, Y.; Ginger, D.; Kalinin, S. V. Semi-supervised learning of images with strong rotational disorder: assembling nanoparticle libraries. *Digital Discovery* 2024, *3* (6), 1213-1220. DOI: 10.1039/D3DD00196B.

(54) Liu, Y.; Huey, B. D.; Ziatdinov, M. A.; Kalinin, S. V. Physical discovery in representation learning via conditioning on prior knowledge. *Journal of Applied Physics* 2024, *136*, 064902. DOI: 10.1063/5.0222403.

(55) Roccapriore, K. M.; Ziatdinov, M.; Cho, S. H.; Hachtel, J. A.; Kalinin, S. V. Predictability of Localized Plasmonic Responses in Nanoparticle Assemblies. *Small* 2021, *17* (21), 2100181. DOI: 10.1002/smll.202100181.

(56) Calvat, M.; Bean, C.; Anjaria, D.; Park, H.; Wang, H.; Vecchio, K.; Stinville, J. Learning Metal Microstructural Heterogeneity through Spatial Mapping of Diffraction Latent Space Features, Ver. 1, *arXiv*, January 2025. DOI: 10.48550/arXiv.2501.18064.

(57) Oommen, V.; Shukla, K.; Goswami, S.; Dingreville, R.; Karniadakis, G. E. Learning two-phase microstructure evolution using neural operators and autoencoder architectures. *Npj Comput Mater* 2022, *8* (1), 190. DOI: 10.1038/s41524-022-00876-7.

(58) Trask, N.; Martinez, C.; Shilt, T.; Walker, E.; Lee, K.; Garland, A.; Adams, D. P.; Curry, J. F.; Dugger, M. T.; Larson, S. R.; et al. Unsupervised physics-informed disentanglement of multimodal materials data. *Materials Today* 2024, *80*, 286-296. DOI: 10.1016/j.mattod.2024.09.005.

(59) Wilson, A. G.; Hu, Z.; Salakhutdinov, R.; Xing, E. P. Deep kernel learning. In *Artificial intelligence and statistics*, 2016; PMLR: pp 370-378.

(60) Valleti, M.; Vasudevan, R. K.; Ziatdinov, M. A.; Kalinin, S. V. Deep kernel methods learn better: from cards to process optimization. *Machine Learning: Science and Technology* 2024, *5* (1), 015012. DOI: 10.1088/2632-2153/ad1a4f.

(61) Adadi, A.; Berrada, M. Peeking Inside the Black-Box: A Survey on Explainable Artificial Intelligence (XAI). *IEEE Access* 2018, *6*, 52138-52160. DOI: 10.1109/ACCESS.2018.2870052.

(62) Mersha, M.; Lam, K.; Wood, J.; AlShami, A. K.; Kalita, J. Explainable artificial intelligence: A survey of needs, techniques, applications, and future direction. *Neurocomputing* 2024, *599*, 128111. DOI: 10.1016/j.neucom.2024.128111.

(63) Hassija, V.; Chamola, V.; Mahapatra, A.; Singal, A.; Goel, D.; Huang, K.; Scardapane, S.; Spinelli, I.; Mahmud, M.; Hussain, A. Interpreting Black-Box Models: A Review on Explainable Artificial Intelligence. *Cognitive Computation* 2024, *16* (1), 45-74. DOI: 10.1007/s12559-023-10179-8.

(64) Zhong, X.; Gallagher, B.; Liu, S.; Kailkhura, B.; Hiszpanski, A.; Han, T. Y.-J. Explainable machine learning in materials science. *Npj Comput Mater* 2022, *8* (1), 204. DOI: 10.1038/s41524-022-00884-7.

(65) Oviedo, F.; Ferres, J. L.; Buonassisi, T.; Butler, K. T. Interpretable and Explainable Machine Learning for Materials Science and Chemistry. *Accounts of Materials Research* 2022, *3* (6), 597-607. DOI: 10.1021/accountsmr.1c00244.

(66) Liu, T.; Barnard, A. S. The emergent role of explainable artificial intelligence in the materials sciences. *Cell Reports Physical Science* 2023, *4* (10), 101630. DOI: 10.1016/j.xcrp.2023.101630.

(67) Mangal, A.; Holm, E. A. A Comparative Study of Feature Selection Methods for Stress Hotspot Classification in Materials. *Integr Mater Manuf I* 2018, *7* (3), 87-95. DOI: 10.1007/s40192-018-0109-8.

(68) Pankajakshan, P.; Sanyal, S.; de Noord, O. E.; Bhattacharya, I.; Bhattacharyya, A.; Waghmare, U. Machine Learning and Statistical Analysis for Materials Science: Stability and Transferability of Fingerprint Descriptors and Chemical Insights. *Chemistry of Materials* 2017, *29* (10), 4190-4201. DOI: 10.1021/acs.chemmater.6b04229.

(69) Liu, Y.; Wu, J.-M.; Avdeev, M.; Shi, S.-Q. Multi-Layer Feature Selection Incorporating Weighted Score-Based Expert Knowledge toward Modeling Materials with Targeted Properties. *Advanced Theory and Simulations* 2020, *3* (2), 1900215. DOI: 10.1002/adts.201900215.

(70) Davies, D. W.; Butler, K. T.; Walsh, A. Data-Driven Discovery of Photoactive Quaternary Oxides Using First-Principles Machine Learning. *Chemistry of Materials* 2019, *31* (18), 7221-7230. DOI: 10.1021/acs.chemmater.9b01519.

(71) Altmann, A.; Toloşi, L.; Sander, O.; Lengauer, T. Permutation importance: a corrected feature importance measure. *Bioinformatics* 2010, *26* (10), 1340-1347. DOI: 10.1093/bioinformatics/btq134.

(72) Simonyan, K.; Vedaldi, A.; Zisserman, A. Deep inside convolutional networks: Visualising image classification models and saliency maps, Ver. 2. *arXiv*, April 2014. DOI: 10.48550/arXiv.1312.6034.

(73) Kondo, R.; Yamakawa, S.; Masuoka, Y.; Tajima, S.; Asahi, R. Microstructure recognition using convolutional neural networks for prediction of ionic conductivity in ceramics. *Acta Materialia* 2017, *141*, 29-38. DOI: 10.1016/j.actamat.2017.09.004.

(74) Oviedo, F.; Ren, Z.; Sun, S.; Settens, C.; Liu, Z.; Hartono, N. T. P.; Ramasamy, S.; DeCost, B. L.; Tian, S. I. P.; Romano, G.; et al. Fast and interpretable classification of small X-ray diffraction datasets using data augmentation and deep neural networks. *Npj Comput Mater* 2019, *5* (1), 60. DOI: 10.1038/s41524-019-0196-x.

(75) Ribeiro, M. T.; Singh, S.; Guestrin, C. "Why Should I Trust You?": Explaining the Predictions of Any Classifier. *Proceedings of 22nd ACM SIGKDD International Conference on Knowledge Discovery and Data Mining*, Association for Computing Machinery, New York, USA, 2016, pp. 1135–1144. DOI: 10.1145/2939672.2939778.

(76) Bordekar, H.; Cersullo, N.; Brysch, M.; Philipp, J.; Hühne, C. eXplainable artificial intelligence for automatic defect detection in additively manufactured parts using CT scan analysis. *Journal of Intelligent Manufacturing* 2025, *36* (2), 957-974. DOI: 10.1007/s10845-023-02272-4.

(77) Iquebal, A. S.; Pandagare, S.; Bukkapatnam, S. Learning acoustic emission signatures from a nanoindentation-based lithography process: Towards rapid microstructure characterization. *Tribology International* 2020, *143*, 106074. DOI: 10.1016/j.triboint.2019.106074.

(78) Lundberg, S. M.; Lee, S.-I. A unified approach to interpreting model predictions, Ver. 2. *arXiv,* November 2017. DOI: 10.48550/arXiv.1705.07874.

(79) Grimberg, H.; Tiwari, V. S.; Tam, B.; Gur-Arie, L.; Gingold, D.; Polachek, L.; Akabayov, B. Machine learning approaches to optimize small-molecule inhibitors for RNA targeting. *Journal of Cheminformatics* 2022, *14* (1), 4. DOI: 10.1186/s13321-022-00583-x.

(80) Rodríguez-Pérez, R.; Bajorath, J. Interpretation of Compound Activity Predictions from Complex Machine Learning Models Using Local Approximations and Shapley Values. *Journal of Medicinal Chemistry* 2020, *63* (16), 8761-8777. DOI: 10.1021/acs.jmedchem.9b01101.

(81) Yazdani, K.; Jordan, D.; Yang, M.; Fullenkamp, C. R.; Calabrese, D. R.; Boer, R.; Hilimire, T.; Allen, T. E. H.; Khan, R. T.; Schneekloth Jr., J. S. Machine Learning Informs RNA-Binding Chemical Space. *Angewandte Chemie International Edition* 2023, *62* (11), e202211358. DOI: 10.1002/anie.202211358.

(82) Wojtuch, A.; Jankowski, R.; Podlewska, S. How can SHAP values help to shape metabolic stability of chemical compounds? *Journal of Cheminformatics* 2021, *13* (1), 74. DOI: 10.1186/s13321-021-00542-y.

(83) Korolev, V. V.; Mitrofanov, A.; Marchenko, E. I.; Eremin, N. N.; Tkachenko, V.; Kalmykov, S. N. Transferable and Extensible Machine Learning-Derived Atomic Charges for Modeling Hybrid Nanoporous Materials. *Chemistry of Materials* 2020, *32* (18), 7822-7831. DOI: 10.1021/acs.chemmater.0c02468.

(84) Jablonka, K. M.; Ongari, D.; Moosavi, S. M.; Smit, B. Big-Data Science in Porous Materials: Materials Genomics and Machine Learning. *Chemical Reviews* 2020, *120* (16), 8066-8129. DOI: 10.1021/acs.chemrev.0c00004.

(85) Guo, S.; Huang, X.; Situ, Y.; Huang, Q.; Guan, K.; Huang, J.; Wang, W.; Bai, X.; Liu, Z.; Wu, Y.; et al. Interpretable Machine-Learning and Big Data Mining to Predict Gas Diffusivity in Metal-Organic Frameworks. *Advanced Science* 2023, *10* (21), 2301461. DOI: 10.1002/advs.202301461.

(86) Hartono, N. T. P.; Thapa, J.; Tiihonen, A.; Oviedo, F.; Batali, C.; Yoo, J. J.; Liu, Z.; Li, R.; Marrón, D. F.; Bawendi, M. G.; et al. How machine learning can help select capping layers to suppress perovskite degradation. *Nature Communications* 2020, *11* (1), 4172. DOI: 10.1038/s41467-020-17945-4.

(87) Wang, J.; Qi, Y.; Zheng, H.; Wang, R.; Bai, S.; Liu, Y.; Liu, Q.; Xiao, J.; Zou, D.; Hou, S. Advancing vapor-deposited perovskite solar cells via machine learning. *Journal of Materials Chemistry A* 2023, *11* (25), 13201-13208. DOI: 10.1039/D3TA00027C.

(88) Bahdanau, D.; Cho, K.; Bengio, Y. Neural machine translation by jointly learning to align and translate, Ver. 7. *arXiv*, May 2016. DOI: 10.48550/arXiv.1409.0473.

(89) Jain, S.; Wallace, B. C. Attention is not explanation. *9th International Joint Conference on Natural Language Processing (EMNLP-IJCNLP), Proceedings of the 2019 EMNLP-IJCNLP,* Hong-Kong, China, November, 2019; Inui K., Jiang J., Ng V., Wan X., Eds.; Association for Computational Linguistics, pp. 11-20. DOI: 10.18653/v1/D19-1002

(90) Serrano, S.; Smith, N. A. Is Attention Interpretable? *57th Annual Meeting of the Association for Computational Linguistics*, *Proceedigs*, Florence, Italy, July, 2019; Korhonen K., Traum D., Màrquez L., Eds.; Association for Computational Linguistics: pp. 2931-2951. DOI: 10.18653/v1/P19-1282.

(91) Brunner, G.; Liu, Y.; Pascual, D.; Richter, O.; Ciaramita, M.; Wattenhofer, R. On identifiability in transformers, Ver. 4. *arXiv*, February 2019. DOI: 10.48550/arXiv.1908.04211.

(92) Louis, S.-Y.; Zhao, Y.; Nasiri, A.; Wang, X.; Song, Y.; Liu, F.; Hu, J. Graph convolutional neural networks with global attention for improved materials property prediction. *Physical Chemistry Chemical Physics* 2020, *22* (32), 18141-18148. DOI: 10.1039/D0CP01474E.

(93) Wang, A. Y.-T.; Kauwe, S. K.; Murdock, R. J.; Sparks, T. D. Compositionally restricted attention-based network for materials property predictions. *Npj Comput Mater* 2021, *7* (1), 77. DOI: 10.1038/s41524-021-00545-1.

(94) Lu, Y.; Sun, Y.; Hou, C.; Li, Z.; Ni, J. Explainable Attention CNN for Predicting Properties of Heusler Alloys. *The Journal of Physical Chemistry C* 2025. DOI: 10.1021/acs.jpcc.5c01624.

(95) Madani, M.; Lacivita, V.; Shin, Y.; Tarakanova, A. Accelerating materials property prediction via a hybrid Transformer Graph framework that leverages four body interactions. *Npj Comput Mater* 2025, *11* (1), 15. DOI: 10.1038/s41524-024-01472-7.

(96) Vaswani, A.; Shazeer, N.; Parmar, N.; Uszkoreit, J.; Jones, L.; Gómez, A. N.; Kaiser, Ł.; Polosukhin, I. Attention Is All You Need. *Adv. Neural Inf. Process. Syst*. 2017, 30, 5998–6008.

(97) Bepler, T.; Zhong, E. D.; Kelley, K.; Brignole, E.; Berger, B. Explicitly Disentangling Image Content from Translation and Rotation with spatial-VAE. Adv. Neural Inf. Process. Syst. 2019, 32, 15409–15419. In Proceedings of the 33rd Neural Information Processing Systems (NeurIPS 2019), Vancouver, Canada, December 8–14, 2019.

(98) Ziatdinov, M.; Kalinin, S. Robust feature disentanglement in imaging data via joint invariant variational autoencoders: from cards to atoms, Ver. 1. *arXiv*, April 2021. DOI: 10.48550/arXiv.2104.10180.

(99) Ziatdinov, M. *pyroVED* 2023. https://github.com/ziatdinovmax/pyroVED (accessed 13.08.2025)

(100) Morioka, H.; Yamada, T.; Tagantsev, A. K.; Ikariyama, R.; Nagasaki, T.; Kurosawa, T.; Funakubo, H. Suppressed polar distortion with enhanced Curie temperature in in-plane 90°-domain structure of a-axis oriented $PbTiO_3$ Film. *Applied Physics Letters* 2015, *106* (4), 042905. DOI: 10.1063/1.4906861.

(101) Liu, Y.; Kelley, K. P.; Funakubo, H.; Kalinin, S. V.; Ziatdinov, M. Exploring Physics of Ferroelectric Domain Walls in Real Time: Deep Learning Enabled Scanning Probe Microscopy. *Advanced Science* 2022, *9* (31), 2203957. DOI: 10.1002/advs.202203957.

(102) Liu, Y.; Kelley, K. P.; Vasudevan, R. K.; Funakubo, H.; Ziatdinov, M. A.; Kalinin, S. V. Experimental discovery of structure–property relationships in ferroelectric materials via active learning. *Nat Mach Intell* 2022, *4* (4), 341-350. DOI: 10.1038/s42256-022-00460-0.

(103) Liu, Y.; Ziatdinov, M.; Kalinin, S. V. Exploring Causal Physical Mechanisms via Non-Gaussian Linear Models and Deep Kernel Learning: Applications for Ferroelectric Domain Structures. *ACS Nano* 2022, *16* (1), 1250-1259. DOI: 10.1021/acsnano.1c09059.

(104) Pratiush, U.; Funakubo, H.; Vasudevan, R.; Kalinin, S. V.; Liu, Y. Scientific Exploration with Expert Knowledge (SEEK) in Autonomous Scanning Probe Microscopy with Active Learning. *Digital Discovery* 2025, *4*, 252-263. DOI: 10.1039/D4DD00277F.

(105) Liu, Y. T.; Vasudevan, R. K.; Kelley, K. P.; Funakubo, H.; Ziatdinov, M.; Kalinin, S. V. Learning the right channel in multimodal imaging: automated experiment in piezoresponse force microscopy. *Npj Comput Mater* 2023, *9* (1), 34. DOI: 10.1038/s41524-023-00985-x.

(106) Liu, Y.; Kelley, K. P.; Vasudevan, R. K.; Zhu, W.; Hayden, J.; Maria, J. P.; Funakubo, H.; Ziatdinov, M. A.; Trolier-McKinstry, S.; Kalinin, S. V. Automated Experiments of Local Non-

Linear Behavior in Ferroelectric Materials. *Small* 2022, *18* (48), e2204130. DOI: 10.1002/smll.202204130.

(107) Jesse, S.; Lee, H. N.; Kalinin, S. V. Quantitative mapping of switching behavior in piezoresponse force microscopy. *Review of Scientific Instruments* 2006, *77* (7), 073702. DOI: 10.1063/1.2214699.

(108) Kusner, M. J.; Paige, B.; Hernández-Lobato, J. M. Grammar Variational Autoencoder. *Proc. Mach. Learn. Res.* 2017, 34th Int. Conf. Mach. Learn., PMLR 70, pp. 1945–1954.

(109) Jørgensen, P. B.; Mesta, M.; Shil, S.; García Lastra, J. M.; Jacobsen, K. W.; Thygesen, K. S.; Schmidt, M. N. Machine learning-based screening of complex molecules for polymer solar cells. *The Journal of Chemical Physics* 2018, *148* (24), 241735. DOI: 10.1063/1.5023563.

(110) Slautin, B. N.; Pratiush, U.; Lupascu, D. C.; Ziatdinov, M. A.; Kalinin, S. V. Integrating Predictive and Generative Capabilities by Latent Space Design via the DKL-VAE Model, Ver. 1. *arXiv*, March 2025. DOI: 10.48550/arXiv.2503.02978.

**Supplementary Information**

**1. Supplementary figures supporting the study**

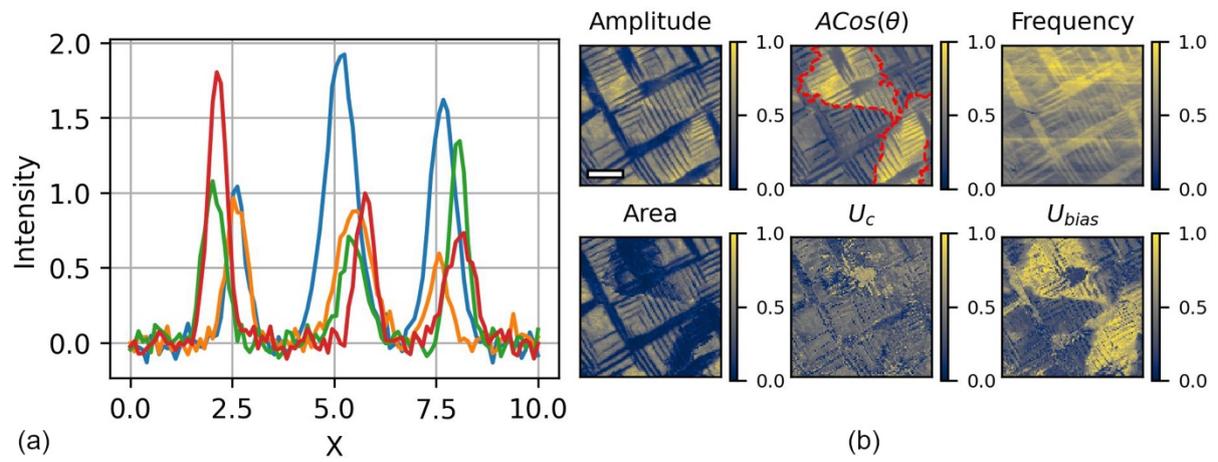

**Figure S1.** (a) Examples of synthetic Gaussian spectra. (b) normalized PFM responses of the local domain structure in a PbTiO$_3$ film. Scale bar: 2 μm. The approximate path of the c–c domain wall is indicated by the red dashed line.

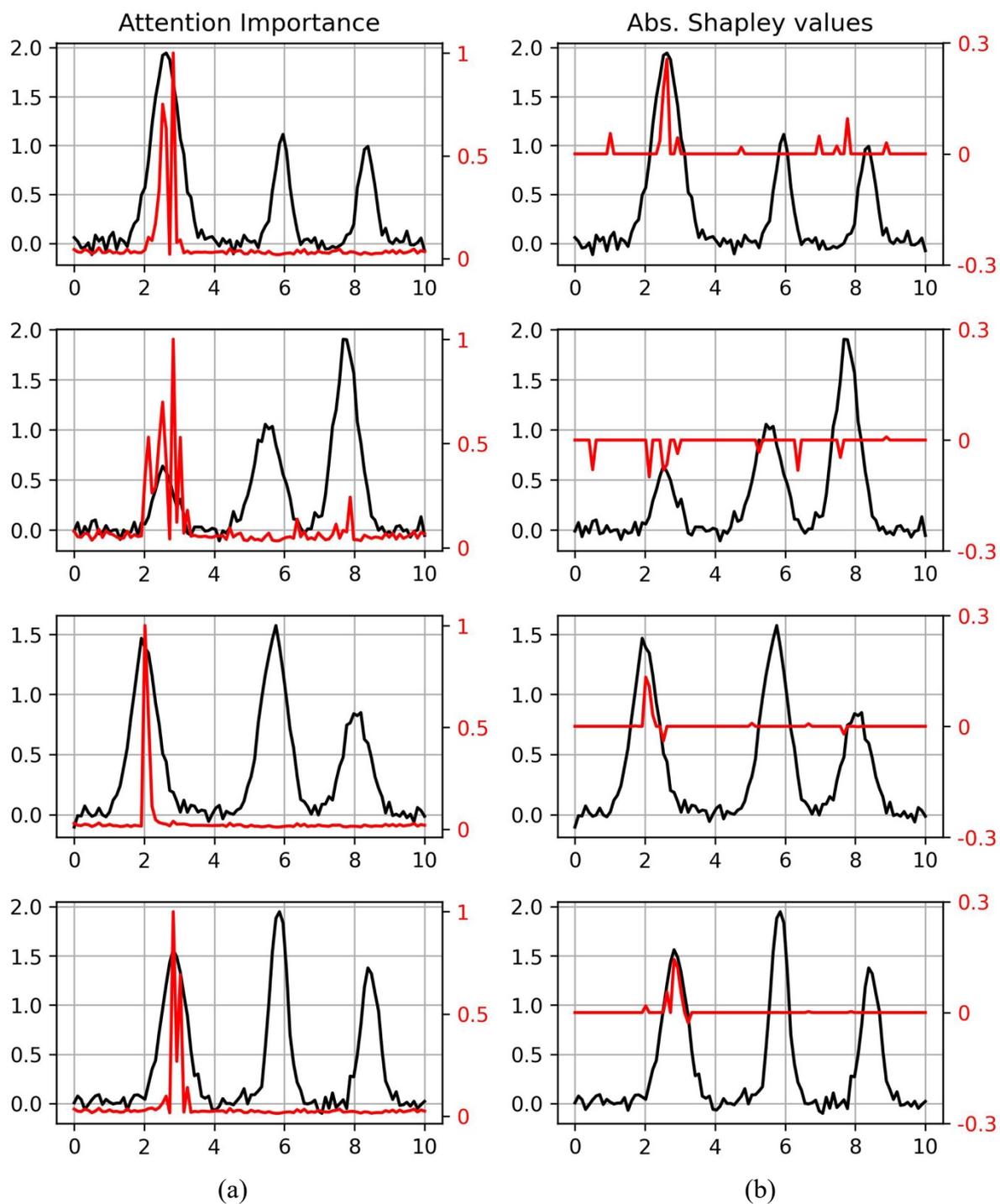

**Figure S2.** Identification of informative regions within the random synthetic test spectra using (a) attention-based importance scores and (b) SHAP values

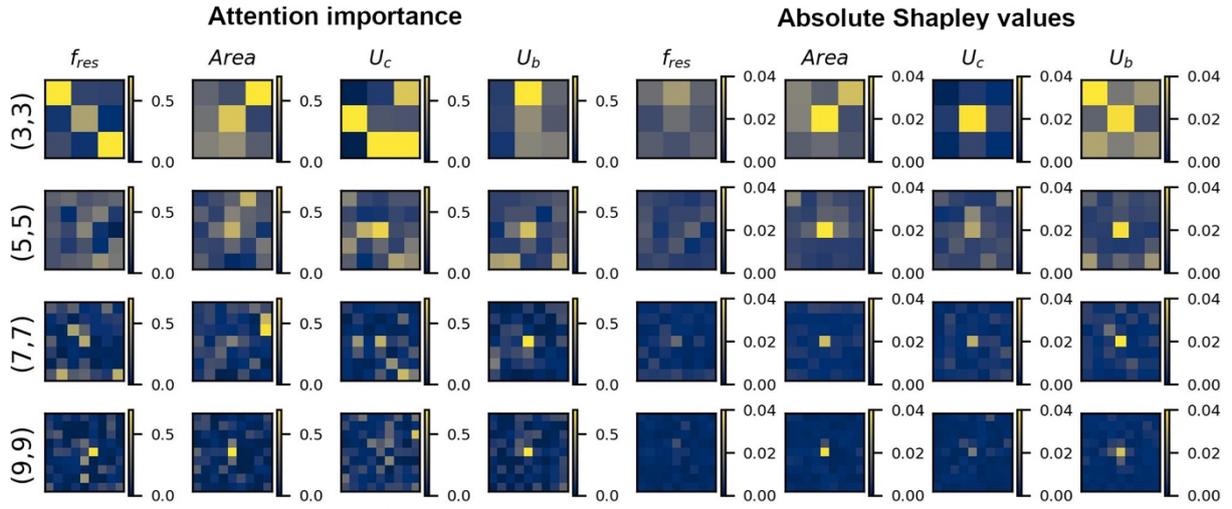

**Figure S3.** Average distributions of (a-d) attention-based importance and (e-h) absolute Shapley values for the prediction of contact resonance frequency, loop area, coercive voltage, and bias voltage, respectively, across structural image patches of different sizes.

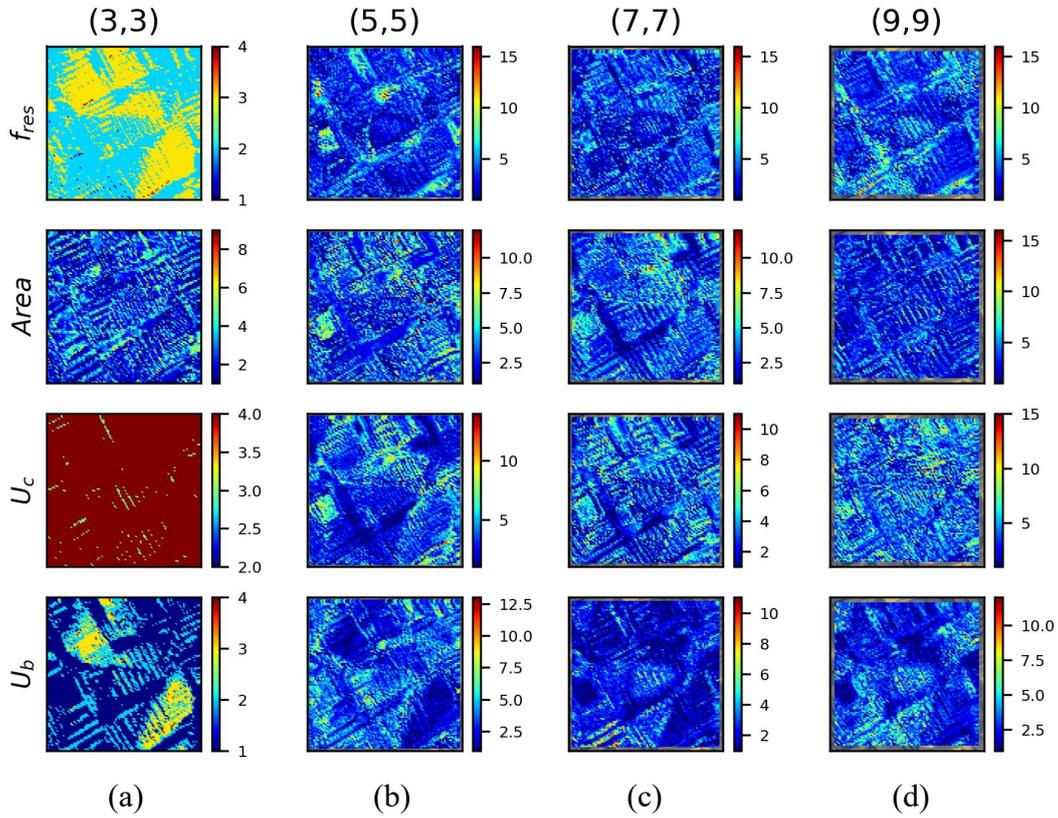

**Figure S4.** Spatial maps showing the number of informative pixels within each image patch, defined as pixels with attention importance values greater than 0.5, for the prediction of resonance frequency, loop area, and coercive voltage. Results are shown for descriptor sizes of (a) 3×3 px, (b) 5×5 px, (c) 7×7 px, and (d) 9×9 px.

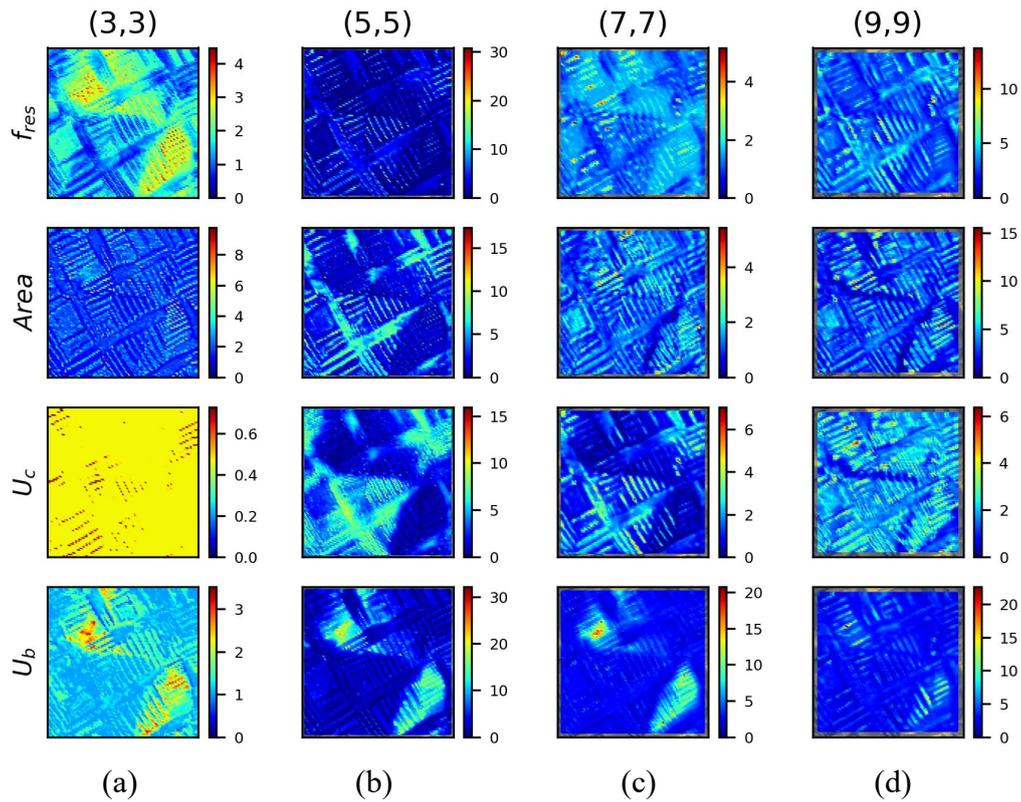

**Figure S5** Spatial maps of the ratio between the attention importance of the central pixels and the average importance of the remaining pixels in the patch for the prediction of resonance frequency, loop area, and coercive voltage. Results are shown for descriptor sizes of (a) 3×3 px, (b) 5×5 px, (c) 7×7 px, and (d) 9×9 px.

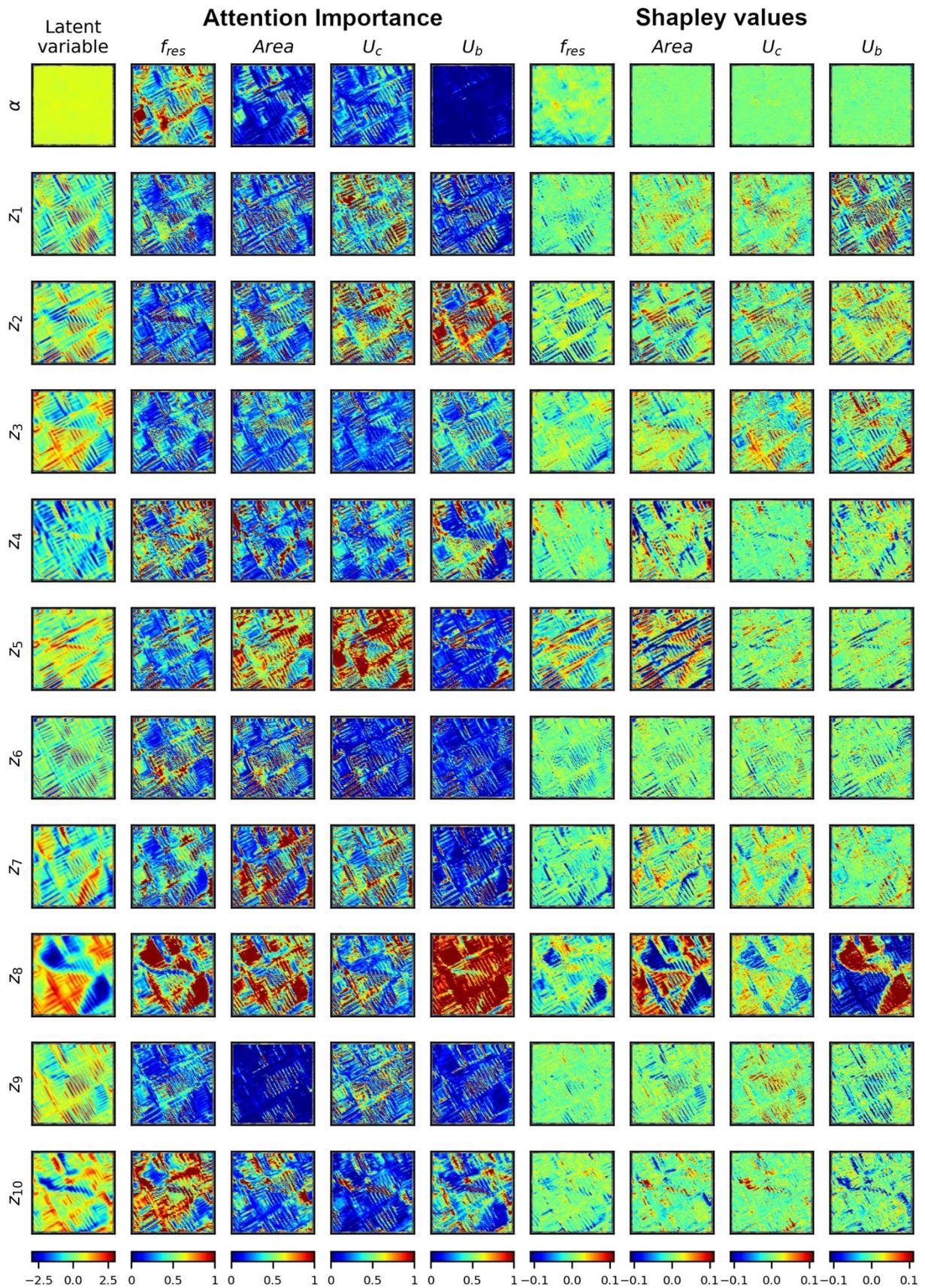

**Figure S6.** Attention-based importance scores and Shapley values projected into real space for latent variables of 10D VAE latent representation.

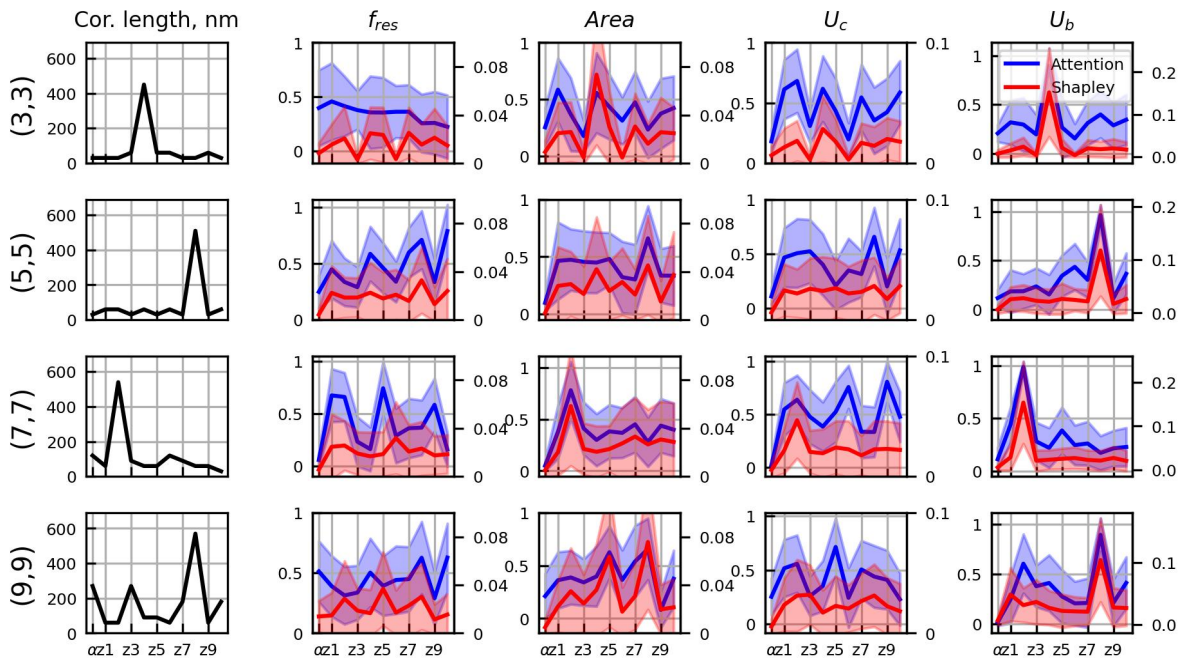

**Figure S7.** Correlation lengths of the rVAE latent variables and their corresponding average attention-based importance and absolute Shapley values for predicting contact resonance frequency, loop area, coercive voltage, and loop bias across different descriptor sizes.

## 2. Prompt and raw output of the LLM-based deep research

Prompt:
"Find relevant studies in materials science and provide a list of key references and short descriptions supporting the following themes:

How structural features govern material functionality, particularly in studies using microscopic techniques such as SEM, SPM, or Raman spectroscopy.

How machine learning models—including supervised learning, autoencoders, or transformer-based models—are applied to model and interpret complex structure–property relationships.

The challenge of explainability in ML-driven modeling, and implementation SHAP (SHapley Additive exPlanations) and other approaches to improve explainability.

Please include both foundational and recent references (2020–2024). Prioritize studies from materials-focused journals."

The raw output of GPT-4o model is available by the link:
https://chatgpt.com/s/dr_6871554257a4819195aec33a5c81dfd9